\documentstyle[12pt,epsf]{article} \hoffset -0.5in \textwidth 6.00in
\textheight
8.5in  \parskip 7pt \openup1.5\jot \parindent=0.5in
\newskip\humongous \humongous=0pt plus 1000pt minus 1000pt
\def\caja{\mathsurround=0pt}
\def\eqalign#1{\,\vcenter{\openup1\jot \caja
        \ialign{\strut \hfil$\displaystyle{##}$&$
        \displaystyle{{}##}$\hfil\crcr#1\crcr}}\,}
\newif\ifdtup

\def\eqright #1\cr{\noalign{\hfill$\displaystyle{{}#1}$}}
\def\eqleft #1\cr{\noalign{\noindent$\displaystyle{{}#1}$\hfill}}

\def\oldreffmt#1{\rlap{[#1]} \hbox to 2\parindent{}}

\def\figfmt#1{\rlap{Figure {#1}} \hbox to 1in{}}

\def\sectioneq{\def\theequation{\thesection.\arabic{equation}}{\let
\holdsection=\section\def\section{\setcounter{equation}{0}\holdsection}}}%

\newcounter{holdequation}

\def\auto{\eqno(\refstepcounter{equation}\theequation)}
\def\begineq #1\endeq{$$ \refstepcounter{equation}\eqalign{#1}\eqno
	(\theequation) $$}
\def\contlimit{\,{\hbox{$\longrightarrow$}\kern-1.8em\lower1ex
\hbox{${\scriptstyle (a\rightarrow0)}$}}\,}
\def\centeron#1#2{{\setbox0=\hbox{#1}\setbox1=\hbox{#2}\ifdim
\wd1>\wd0\kern.5\wd1\kern-.5\wd0\fi
\copy0\kern-.5\wd0\kern-.5\wd1\copy1\ifdim\wd0>\wd1
\kern.5\wd0\kern-.5\wd1\fi}}
\def\centerover#1#2{\centeron{#1}{\setbox0=\hbox{#1}\setbox
1=\hbox{#2}\raise\ht0\hbox{\raise\dp1\hbox{\copy1}}}}
\def\centerunder#1#2{\centeron{#1}{\setbox0=\hbox{#1}\setbox
1=\hbox{#2}\lower\dp0\hbox{\lower\ht1\hbox{\copy1}}}}
\def\lsim{\;\centeron{\raise.35ex\hbox{$<$}}{\lower.65ex\hbox
{$\sim$}}\;}
\def\gsim{\;\centeron{\raise.35ex\hbox{$>$}}{\lower.65ex\hbox
{$\sim$}}\;}

\def\super#1{\ifmmode \hbox{\textsuper{#1}}\else\textsuper{#1}\fi}
\def\textsuper#1{\newcount\holdspacefactor\holdspacefactor=\spacefactor
$^{#1}$\spacefactor=\holdspacefactor}

\def\getcite#1,{\advance\citenumber by1
\def\getcitearg{#1}\def\lastarg{@}
\ifnum\citenumber=1
\ref{#1}\let\next=\getcite\else\ifx\getcitearg\lastarg\let\next=\relax
\else ,\ref{#1}\let\next=\getcite\fi\fi\next}

\def\pom{{\rm P\kern -0.53em\llap I\,}}
\def\spom{{\rm P\kern -0.36em\llap \small I\,}}
\def\sspom{{\rm P\kern -0.33em\llap \footnotesize I\,}}

\relax

\newskip\humongous \humongous=0pt plus 1000pt minus 1000pt
\def\caja{\mathsurround=0pt}
\def\eqalign#1{\,\vcenter{\openup1\jot \caja
        \ialign{\strut \hfil$\displaystyle{##}$&$
        \displaystyle{{}##}$\hfil\crcr#1\crcr}}\,}
\newif\ifdtup

\def\eqright #1\cr{\noalign{\hfill$\displaystyle{{}#1}$}}
\def\eqleft #1\cr{\noalign{\noindent$\displaystyle{{}#1}$\hfill}}

\def\oldreffmt#1{\rlap{[#1]} \hbox to 2\parindent{}}

\def\figfmt#1{\rlap{Figure {#1}} \hbox to 1in{}}

\def\auto{\eqno(\refstepcounter{equation}\theequation)}
\def\begineq #1\endeq{$$ \refstepcounter{equation}\eqalign{#1}\eqno
	(\theequation) $$}
\def\contlimit{\,{\hbox{$\longrightarrow$}\kern-1.8em\lower1ex
\hbox{${\scriptstyle (a\rightarrow0)}$}}\,}
\def\centeron#1#2{{\setbox0=\hbox{#1}\setbox1=\hbox{#2}\ifdim
\wd1>\wd0\kern.5\wd1\kern-.5\wd0\fi
\copy0\kern-.5\wd0\kern-.5\wd1\copy1\ifdim\wd0>\wd1
\kern.5\wd0\kern-.5\wd1\fi}}
\def\centerover#1#2{\centeron{#1}{\setbox0=\hbox{#1}\setbox
1=\hbox{#2}\raise\ht0\hbox{\raise\dp1\hbox{\copy1}}}}
\def\centerunder#1#2{\centeron{#1}{\setbox0=\hbox{#1}\setbox
1=\hbox{#2}\lower\dp0\hbox{\lower\ht1\hbox{\copy1}}}}
\def\lsim{\;\centeron{\raise.35ex\hbox{$<$}}{\lower.65ex\hbox
{$\sim$}}\;}
\def\gsim{\;\centeron{\raise.35ex\hbox{$>$}}{\lower.65ex\hbox
{$\sim$}}\;}

\def\super#1{\ifmmode \hbox{\textsuper{#1}}\else\textsuper{#1}\fi}
\def\textsuper#1{\newcount\holdspacefactor\holdspacefactor=\spacefactor
$^{#1}$\spacefactor=\holdspacefactor}

\def\getcite#1,{\advance\citenumber by1
\ifnum\citenumber=1
\ref{#1}\let\next=\getcite\else\ifx#1@\let\next=\relax
\else ,\ref{#1}\let\next=\getcite\fi\fi\next}

\def\upon #1/#2 {{\textstyle{#1\over #2}}}
\relax
\renewcommand{\thefootnote}{\fnsymbol{footnote}}

\def\mainhead#1{\setcounter{equation}{0}\addtocounter{section}{1}
  \vbox{\begin{center}\large\bf #1\end{center}}\nobreak\par}
\sectioneq
\def\subhead#1{\bigskip\vbox{\noindent\bf #1}\nobreak\par}

\def\til#1{\centeron{\hbox{$#1$}}{\lower 2ex\hbox{$\char'176$}}}
\def\tild#1{\centeron{\hbox{$\,#1$}}{\lower 2.5ex\hbox{$\char'176$}}}
\def\sumtil{\centeron{\hbox{$\displaystyle\sum$}}{\lower
-1.5ex\hbox{$\widetilde{\phantom{xx}}$}}}

\def\pom{{\rm P\kern -0.53em\llap I\,}}
\def\spom{{\rm P\kern -0.36em\llap \small I\,}}
\def\sspom{{\rm P\kern -0.33em\llap \footnotesize I\,}}

\newcommand{\bit}{\begin{itemize}}
\newcommand{\eit}{\end{itemize}}

\newcommand{\beq}{\begin{equation}}
\newcommand{\eeq}{\end{equation}}
\newcommand{\beqa}{\begin{eqnarray}}
\newcommand{\eeqa}{\end{eqnarray}}

\begin{document}
\begin{titlepage}
\rightline{\vbox{\halign{&#\hfil\cr
&ANL-HEP-PR-95-57\cr
&\today\cr}}}
\vspace{0.25in}

\begin{center}

{\large\bf
Buchm\"uller Scaling, the QCD Pomeron, and Tevatron Diffractive Hard
Scattering}

\medskip

Alan R. White

\footnote{Work supported by the U.S. Department of
Energy, Division of High Energy Physics, Contract\newline W-31-109-ENG-38}
\\ \smallskip
High Energy Physics Division, Argonne National Laboratory, Argonne, IL
60439.\\

\end{center}

\begin{abstract}

We discuss the observed scaling of the small-x diffractive and total
deep-inelastic structure functions at HERA. We argue that the parton
interpretation of Buchm\"uller and Hebecker can be understood within QCD as
the appearance of the Pomeron in a Super-Critical phase. In diffractive hard
scattering, the Pomeron appears as reggeized gluon exchange in a
color-compensating background field. The formalism can also be applied to
diffractive $W$ production at the Tevatron. If the scaling is a true
asymptotic property then it should anticipate the appearance of a further
massive sector of QCD and associated asymptotic Critical Pomeron behaviour.

\end{abstract}

\renewcommand{\thefootnote}{\arabic{footnote}} \end{titlepage}

\mainhead{1. INTRODUCTION}

Presently diffractive physics is thought to be non-perturbative and
uncalculable
within QCD. It is hoped, however, that in limited kinematic circumstances
involving small-x and large $Q^2$, semi-perturbative BFKL Pomeron
calculations\cite{lip} will be applicable. Unfortunately, the BFKL Pomeron
is a complicated multi-gluon phenomenon which, from the experimental point
of view, has proven difficult to isolate. In contrast, Buchm\"uller\cite{buc}
has noticed a remarkably simple scaling relation between the diffractive and
full deep-inelastic cross-sections measured at small-x at HERA. Essentially,
both cross-sections have the same $Q^2$ and $x$-dependence\cite{hera}. This
is illustrated in Fig.~1.1 using the ZEUS data but the H1 data give the same
result.

\leavevmode
\epsfxsize=5in
\epsffile{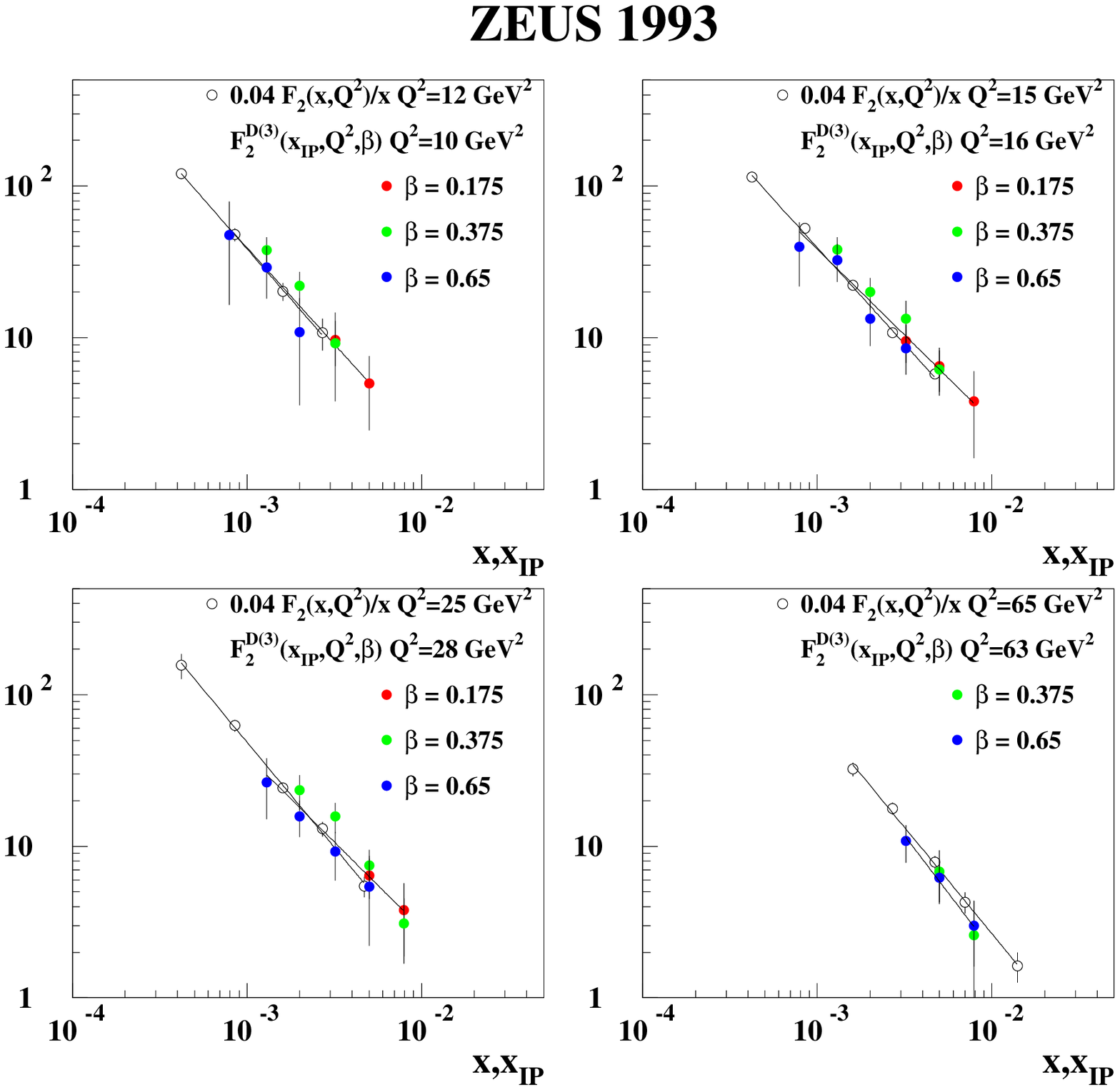}

\begin{center}
Fig.~1.1 Scaling of the Diffractive and Full Deep-Inelastic
Structure Functions - the notation is that of Section 2.
\end{center}

The scaling of Fig.~1.1 suggests an extraordinary simplicity for the Pomeron at
large $Q^2$ - in a sense (that we describe below) {\it it acts like a single
gluon}. Our purpose in this paper is both to explain this simplicity in the
context of our previous study of the QCD Pomeron\cite{arw1,arw2}, and to
expand the picture to hadronic diffractive hard scattering. (It is many
years since we first pointed out\cite{arw} that, in a ``Super-Critical
Phase'', the QCD Pomeron would appear like a single {\it reggeized} gluon.)
Conventionally, the deep-inelastic diffractive cross-section is believed to
contain both non-perturbative ``soft Pomeron'' physics and BFKL
contributions, while the full cross-section is usually evaluated
perturbatively (with BFKL evolution included). Even if the diffractive
cross-section were completely given\cite{bar} by BFKL, no simple scaling
relation between the two cross-sections would be anticipated.

There may be various models that give the proposed scaling behavior either
directly or approximately\cite{bj}. Also it may appear less significant
experimentally as the accuracy of the data improves. However, in a recent
paper\cite{bh}  Buchm\"uller and Hebecker (BH) have made an attractive proposal
that we would like to take very seriously. They suggest that the scaling is
a direct consequence of the rapidity gap (diffractive) events originating
from the same small-x parton scattering process as the normal deep-inelastic
events, i.e. the photon-gluon fusion process of Fig.~1.2.

\begin{center}

\epsffile{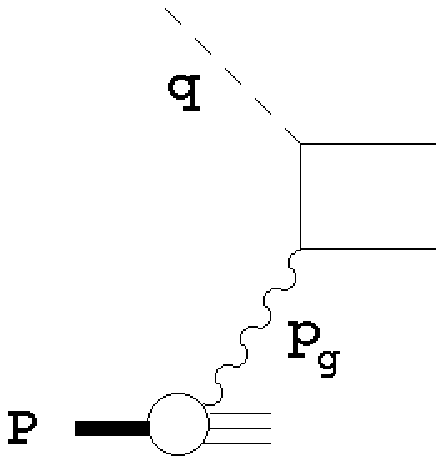}

Fig.~1.2 Photon-Gluon Fusion

\end{center}

\noindent More importantly, BH also suppose that the only subsequent
interaction is very simple. It is pictured as a rescattering in a
``classical'' color field, formed by wee-partons in the proton, which
randomly rotates the color of the quark-antiquark pair. The probability that
the quark final state is color neutral (and so produces the diffractively
excited final state) is then $1/9$ th. This successfully predicts the
experimental ratio of the diffractive cross-section to the total
cross-section and the scaling follows almost trivially. The Pomeron that is
exchanged is clearly a simple gluon - plus only a ``color correction''.

In this paper we will discuss how the, at first sight paradoxical, simplicity
of the BH picture can be understood directly within QCD if the Pomeron is in a
Super-Critical phase of Reggeon Field Theory. (Although we will replace $1/9$
th by $1/8$ th!) Indeed the basis for our analysis\cite{arw1} of the Pomeron
in QCD is to initially identify this phase in a gauge theory framework. Part
of the purpose of this paper will be to discuss how the first approximation,
in which the Pomeron is a single (reggeized) gluon, is exposed by the
deep-inelastic limit.

There must, of course, be a compensation for the exchanged color of the
gluon. In our work there is a ``reggeon condensate'' which plays a role very
similar to the classical wee-parton field envisaged in the BH picture. The
reggeon condensate arises from infra-red divergences of Regge limit transverse
momentum diagrams involving massless quarks. However, we believe it should
also have an interpretation as a classical field related to the
regularization of (multi-)quark operators defining physical states at
infinite momentum. The massless fermion axial U(1) anomaly is crucial for
this and we suspect that the Super-Critical Pomeron can be associated with a
generalized ``winding-number condensate'' arising, essentially, from
instanton interactions. In analogy with the role of the anomaly in the
Schwinger model\cite{man}, this condensate can be viewed as producing
confinement and chiral symmetry breaking {\it at high-energy}. It is not an
SU(3) invariant but rather is an octet (resulting from the restriction of
SU(3) instantons to an SU(2) subgroup). Therefore, in the Super-Critical
phase, an infinite momentum  hadron is an octet quark state in the presence
of the condensate. Gauge-invariance requires an averaging over all SU(3)
directions. Nevertheless, within this averaging, it is possible for single
gluon exchange between the quarks to be accompanied by an (instanton)
interaction which simply realigns the direction of the condensate. This is
the underlying idea in our version of the BH picture. For the Pomeron, the
net color exchanged, by the combination of a gluon and an instanton must be
zero. In a general perturbative gluon interaction this need not be the case
and the result is a factor of 1/8 th.

We will see that as the Pomeron, in a sense, becomes simpler at the partonic
level, it also becomes a more ``non-perturbative'' phenomenon in which
confinement, and not just the requirement of color zero exchange, plays an
essential role. The Super-Critical condensate is a confinement
mechanism which is operative at the scale of the hard scattering. It is
related to, but goes much further than, the color zero cancelation of
infra-red divergences. We find that if the Pomeron is to couple to a hard
scattering process, the parton final state created must be converted to a
true hadronic state by the condensate. This is straightforwardly the case
for the quark-antiquark state of Fig.~1.2, but in general is a strong
constraint on the possible parton states involved. A particular consequence
is that the Ingelman-Schlein assumption\cite{ing} of a universal hadronic
parton distribution function for the Pomeron is not valid. Indeed, when our
formalism is extended to hadronic diffractive $W$ and jet production, the
confinement requirement actually resolves what appear to be puzzling
features of the Tevatron data when the Ingelmam-Schlein formalism is
used\cite{cdf}. Since it is a simple Regge pole, there is also an obvious
simplification of the theoretical factorization properties of the hard
Pomeron within the Super-Critical framework. In general it seems that a
simpler and more unified picture of diffractive physics and the Pomeron may
emerge if we develop the BH picture by exploiting our formalism.

The Super-Critical Pomeron is a (transverse momentum) cut-off dependent
construction and as the effective cut-off increases with energy, it will
eventually disappear unless the true asymptotic behavior of the theory is
given by the Critical Pomeron\cite{cri}. Therefore deep-inelastic
diffractive scaling, or ``Buchm\"uller scaling'', can occur as an {\it exact
asymptotic property} only in this very special circumstance. In fact we have
argued for a long time that the Critical Pomeron may be the only way to
obtain self-consistent unitary asymptotic behavior in the Regge limit and
that a necessary higher-mass quark sector\cite{arw1} should therefore exist.
It may seem surprising that an asymptopic Pomeron property could be directly
related to deep-inelastic scaling at the relatively low energy scale of
HERA. However, if we take the BH picture seriously and ask under what
circumstances it could provide the underlying physical explanation of an
asymptotic scaling property, this is what we are led to.

Unfortunately many of the concepts involved in our work are either
unfamiliar or complicated and technically it is certainly incomplete. As a
result we will give only an outline in this paper that concentrates on
illustration of the ideas involved. Our hope is that a connection with the
BH work will help demonstrate the underlying simplicity and physical
relevance of the picture of the Pomeron that we have arrived at, and also
lead to further applications.

We will break our discussion down into relatively small Sections, beginning
with a summary of the BH paper in Section 2. In Section 3 we will, very
briefly, review what conventional QCD/Pomeron physics has to say about
deep-inelastic diffractive scaling. Three Sections follow that are
self-contained descriptions of crucial elements in our argument. In Section
4 we discuss the conditions under which asymptotic freedom can be combined
with gluon mass generation via the Higgs mechanism. In Section 5 we describe
the analysis of Regge limit infra-red divergences and the resulting Pomeron
when the gauge symmetry of QCD is broken to SU(2), together with the
conditions for the Critical Pomeron to occur. Section 6 contains a very
short description of the Super-Critical Pomeron together with a discussion
of the role of the cut-off in the phase-transition. This allows us to
discuss how the Super-Critical phase occurs in conventional QCD. We return
to deep-inelastic scattering in Section 7 and discuss how the BH
picture is realised in QCD. Section 8 contains our discussion of diffractive
hard scattering in hadron-hadron scattering. Section 9 is a final comment on
the possible new physics at higher-energy scales that is implied if
``Buchmuller scaling'' is a true asymptotic property.

\mainhead{2. THE BUCHM\"ULLER AND HEBECKER FORMALISM}

We begin with a very brief summary of the BH paper. Starting from the
quark-antiquark pair production of Fig.~1.2, the contribution to
the inclusive structure function $F_2(x,Q^2)$ is (in the ``massive gluon''
scheme\cite{field})
\beq\label{f2g}
\Delta^{(g)}F_2(x,Q^2)= x{\alpha_s\over 2\pi} \sum_q e_q^2
\int_x^1{d\xi\over \xi}g(\xi) F(\beta,Q^2)\ .
\eeq
where $g(x)$ is the gluon density and, if we define momenta as in Fig.~1.2 so
that
\beq
 Q^2=-q^2\ ,\ ~~~x={Q^2\over 2 P\cdot q}\ ,~~~~M^2=(q+p_g)^2\ .
\eeq
and write $\vec{p}_g = x_{\spom} \vec{P}$ so that when
$-p_g^2 = m_g^2 ~\ll~ Q^2,~M^2$
\beq
\beta \equiv {Q^2\over Q^2+M^2} \simeq {x\over x_{\spom}}\ .
\eeq
we have
\beq\label{F}
F(\beta,Q^2)~=~\left((\beta^2+(1-\beta)^2)\ln{Q^2\over m_g^2 \beta^2}
- 2 + 6\beta(1-\beta) \right).
\eeq
Assuming that $g(x)$ is power-behaved at small $x$ leads to a  simple
approximation for $F_2$, i.e.
\beq\label{fit}
F_2(x,Q^2) \simeq {\alpha_s\over 3\pi}\sum_q e_q^2 x g(x)
\left({2\over 3} + \ln{Q^2\over m_g^2}\right)\ .
\eeq
which (with a simple parameterization for $g(x)$ and with $m_g~\sim~1$ GeV)
fits the data very well .

In fact (\ref{fit}) works much better than it should since, in principle, it
only describes that part of the cross-section in which the final state contains
two (quark) jets. That kinematically either the quark or the antiquark is
likely to be relatively soft corresponds, of course, to the experimental
feature that the bulk of the cross-section contains only one jet. If the soft
quark (anti-quark) is produced by the photon, the corresponding contribution
to the photon structure function would simply be given by the appropriate
part of Fig.~1.2. A similar remark would apply when the soft quark
(anti-quark) is produced by the gluon if this carries large enough momentum
transfer. As the momentum transfer becomes smaller, we might expect
the single gluon to be replaced by a (multi-gluon) Pomeron. If the gluon in
Fig.~1.2 merges smoothly into the Pomeron, as we are about to describe, this
may
actually be connected with the ability of (\ref{fit}) to approximate far
more of the cross-section than it naively should. This point will be
important later.

The assumption that diffractive deep-inelastic scattering is described by
the same process as pictured in Fig.~1.2, except that the quark-antiquark pair
evolves into a color singlet state with probability $1/9$, predicts\cite{bh}
that
\beq\label{f2d}
F_2^D(x_{\spom},Q^2,\beta)\simeq {1\over 9}{\alpha_s\over 2\pi} \sum_q e_q^2
g(x_{\spom})\bar{F}_2^D(\beta,Q^2)\ ,
\eeq
where
\beq\label{fpom}
\bar{F}_2^D(\beta,Q^2) = \beta F(\beta,Q^2)
\eeq
Since both $g(x_{\spom})$ and $m_g$ have been determined by the fit to the
inclusive structure function $F_2$, the diffractive structure function is
unambiguously predicted, including its normalization. For $\beta$ between
$0.2$ and $0.6$, $\bar{F}_2^D(\beta,Q^2)$ varys rather slowly with $\beta$.
As a result, in this interval,
\beq\label{scaling}
F_2^D(x_{\spom},Q^2,\beta)\simeq {C\over x_{\spom} } F_2(x_{\spom},Q^2)\ ,
\eeq
where $C\simeq 0.04$, independent of $Q^2$. This is the scaling relation,
first suggested by Buchm\"uller\cite{buc}, which is tested in Fig.~1.1.
($F_2^{D(3)}$ should be identified with $F_2^D$). We conclude that the BH
model provides a very successful description of the properties of the large
rapidity gap deep-inelastic events when $x$ is small and $Q^2$ and $M^2$ are
large - so that the cross-section is relatively {\it insensitive to the
gluon mass $m_g$.}

Note, however, that the gluon mass has played two essential roles. Firstly
it has provided the scale for the $Q^2$ scaling violation. Secondly,
without a gluon mass, higher-order corrections which reggeize the gluon
exchange producing the diffractive cross-section would be infra-red divergent
- providing a strong exponential suppression of this exchange.

\mainhead{3. CONVENTIONAL QCD/POMERON PHYSICS }

To put the following discussion in context, we briefly outline what
conventional QCD/Pomeron physics has to say about the diffractive
cross-section.
Suppose first that large $Q^2$ produces large transverse momentum exchange
across the rapidity gap, so that perturbation theory may be applicable. The
simplest color-zero ``perturbative Pomeron'', which avoids the exponential
suppression by infra-red divergences, is two gluons. To calculate scaling
violations, with a conventional factorization scale replacing the gluon
mass, a factorization theorem must be proved. Attempts to derive such a
theorem have led to various problems\cite{fac,bs} and indeed the
factorization properties are different in deep-inelastic and hadron
scattering. (This actually reflects the fact that the ``Pomeron'' involved is
not a Regge pole.) To the extent that there is factorization, a new parton
distribution appears\cite{bs}, not the gluon distribution of the BH picture.

If the issue of the $Q^2$ scaling violations is ignored, and one looks
instead at the small-x dependence, higher-order corrections lead to the
reggeization of both gluons, together with reggeon interactions, and the
BFKL Pomeron is formed across the rapidity gap. This is illustrated in
Fig.~3.1 and discussed at length in \cite{bar}.

\begin{center}

\epsffile{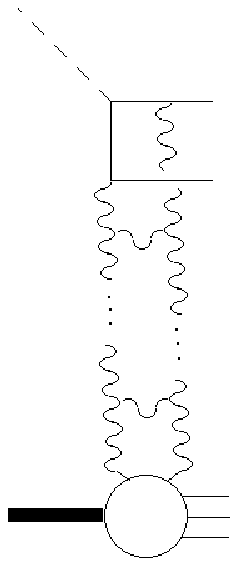}

Fig.~3.1 BFKL Pomeron Production of a Rapidity Gap

\end{center}

\noindent  If we compare with the calculation of the total deep-inelastic
cross-section via BFKL then higher-order corrections, producing the
square of the gluon distribution are involved. This implies that the
diffractive structure function should rise faster at small-x than the full
structure function.

It is anticipated, of course, that the bulk of the diffractive cross-section
involves relatively small momentum transfer across the rapidity gap, even
when $Q^2$ and $M^2$ are large. We then expect the ``soft Pomeron'' to be
involved. The conventional (Ingelman and Schlein\cite{ing}) assumption is
that the large $Q^2$ is absorbed by a parton interaction described by a
Pomeron structure function. Since non-perturbative physics is invoked there
is no reason to expect any very simple relation with respect to the
perturbatively calculated total cross-section.

Note that, at any fixed $Q^2$, the asymptotic small-x behaviour of the total
cross-section is, of course, given by the full non-perturbative Pomeron.
In this case the scaling behavior we are looking for requires that the
$x$-dependence (i.e the cross-section increase with energy) be the same for
the one Pomeron exchange that gives the total cross-section and the
triple-Pomeron interaction that gives the high-mass rapidity-gap
cross-section. This is a consistency condition for the asymptotic Pomeron
that is satisfied\cite{abbs}, so far as is known, only by the Critical
Pomeron. This makes it plausible that the appearance of combined
deep-inelastic and diffractive scaling may actually be the short-distance
precursor of Critical Pomeron asymptotic behavior.

\mainhead{4. A GLUON MASS FROM SYMMETRY BREAKING}

A simple  prerequisite for the BH calculation would be that a gluon mass can
be smoothly introduced into the calculation of scaling violations without
destroying asymptotic freedom. To introduce a gluon mass in a
gauge-invariant way requires the use of the Higgs mechanism. A-priori, we
then expect to lose contact with confining, asymptotically-free, unbroken
QCD. However, as observed in the original asymptotic freedom paper of Gross
and Wilzcek\cite{gw}, we can break the symmetry from $SU(3)$ to $SU(2)$ by
using a single fundamental representation Higgs scalar. In this special
case, asymptotic freedom of both the gauge coupling and the scalar
self-coupling can be preserved {\it if enough fermions are present}, as we
briefly illustrate.

If $g(t)$ and $h(t)$ are respectively the gauge and scalar scale-dependent
couplings, then
$$
\eqalign{ {{dg} \over {dt}} =  -{1\over 2}b_0t^3 + \cdots}
\auto\label{3.2}
$$
where
$$
b_0 = {1\over {8\pi^2}} \left[11 -{4\over 3}S_3(f)-{1\over 6}\right]
\auto
$$
The 1/6 is due to the triplet scalar and $S_3(f)$ depends on the number (and
representation) of the quarks present. Similarly
$$
\eqalign{{{dh}\over {dt}} = Ah^2 + Bg^2 + Cg^4 + \cdots }
\auto\label{3.5}
$$
where
$$
A = {7\over {8\pi^2}},\ B = -{1\over {\pi^2}}~~~ and~~
C = {{13}\over {48\pi^2}}.
\auto
$$
Requiring that $h ~\sim~g^2~\rightarrow 0$ consistently gives a stability
equation which requires that $b_0$ be very small i.e.
$$
{5\over {24}} > 8\pi^2 b_0
\auto\label{nf}
$$
for a solution.

(\ref{nf}) is satisfied if, and only if, the maximum number of fermions
consistent with asymptotic freedom are present, i.e. $N_f=N_f^{max}$. In
order to relax this condition we will need to introduce Reggeon Field Theory
explicitly into our discussion, as we will do shortly. This will enable us
to understand the general significance, for the Pomeron, of breaking the
gauge symmetry from SU(3) to SU(2), independently of the number of fermions
present. For the moment we note that the large number of fermions is a
prerequisite only if we insist on combining asymptotic freedom with the
symmetry breaking generating the gluon mass. Of course, the additional
fermions can be present at a higher mass scale. The effective theory
obtained by integrating them out will be asymptotically free and the  gluon
mass can be smoothly generated as we have discussed. Note also that as the
gluon mass is taken to zero, the scalar mass can be simultaneously sent to
infinity so that it decouples from the theory. Because it is in the
fundamental representation the complimentarity principle applies to this
decoupling, i.e. there is a smooth relation between the ``Higgs region'' of
parameter space and the ``confinement region''. A related property is that
a gauge-invariant description of the ``symmetry breaking'' can be given. A
full discussion can be found in \cite{arw1}.

The spectrum of gluons when the symmetry is broken is important for our
discussion. Under the remaining SU(2) symmetry we have

\begin{itemize}

\item{ one massless triplet}

\item{two massive doublets}

\item{one massive singlet}

\end{itemize}

\noindent giving, of course, eight in total. The massless triplet is
responsible for the remaining SU(2) gauge symmetry while the singlet is the
essential feature for our purposes.

\mainhead{5. INFRA-RED DIVERGENCES AND THE WEE-PARTON CONDENSATE}

An (extremely) non-trivial part of our discussion is the emergence of a
reggeon, or ``wee-parton'',  condensate when the gauge symmetry is broken as
discussed in the last Section. It underlies our understanding of the
Super-Critical Pomeron and is the basis for our argument that a
semi-classical field can indeed be responsible for color compensation of
parton scattering, as in the BH picture. The theoretical problem of
determining conditions under which the wee partons in a hadron can be
described by such a field may appear to be novel. However, it first arose
nearly twenty years ago as the problem of making sense of the notion of a
``Pomeron condensate" defining the Super-Critical Reggeon Field
Theory\cite{arw2} that we discuss in the next Section.

According to our analysis\cite{arw1}, if the gauge symmetry is broken from
SU(3) to SU(2) then, {\it in the diffractive scattering regime,} all the
infra-red divergences of reggeon diagrams due to the remaining massless
gluons can be absorbed in a stable ``reggeon condensate''. If this
symmetry-breaking is to be a smooth effect at all transverse momenta then
asymptotic freedom must be preserved as discussed in the last Section. {\it If
$N_f<N_f^{max}$}, so that this is not the case, {\it a transverse momentum
cut-off must be introduced as part of the analysis}. In the next Section
we will see that this cut-off becomes an important dynamical variable.

As is made clear, \cite{arw1} is at best an outline of a full analysis. We
summarize the results first and then elaborate briefly on their origin.

\begin{itemize}

\item{The presence of light quarks (initially taken to be massless) produces
Regge limit infra-red ``anomalous'' interactions which generate a wide
ranging exponentiation of infra-red divergences. In SU(2) gauge theory an
overall divergence remains which can be absorbed, as a ``condensate'', into
the scattering states. The divergence selects out the confinement plus
chiral symmetry breaking spectrum of quark hadronic states and is the only
high-energy remnant of multi-gluon exchange - there is no Pomeron.  }

\item{For SU(3) gauge symmetry broken to SU(2), a Regge pole Pomeron is
formed by the SU(2) singlet gluon, which is reggeized, in the background of
the condensate and can be identified with the Super-Critical Pomeron of
Reggeon Field Theory. If the gauge symmetry is restored to SU(3), the
condensate disappears and, straightforwardly when $N_f=N_f^{max}$, the
Critical Pomeron is obtained.}

\end{itemize}

(Although it does not appear in \cite{arw1}, we should note that the
even signature symmetric octet two gluon reggeon discovered in \cite{bar}
becomes exchange degenerate with the reggeized gluon as the SU(3) symmetry
is restored. It also contributes to the Pomeron and is a crucial component
at small momentum transfer. For simplicity we shall make only brief
references to this in the following since it is irrelevant to our essential
purpose - the discussion of diffractive hard scattering.)

It is, of course, well-known that a cloud of soft photons must be
introduced in QED to define gauge-invariant electron states and this manifests
itself in high-energy scattering as the divergent Coulomb phase. In a
non-abelian theory, with massless quarks, our analysis implies that this
divergence problem is dominated by effects related to topological gauge fields
and the massless fermion anomaly equation i.e.
$$
\partial_\mu j^\mu_5 ~=~ -N_f F\tilde{F}~/16\pi^2 ~=~
-N_f \partial_\mu K^\mu ~/16\pi^2
\auto\label{ano}
$$
$j^{\mu}_5$ is the color singlet U(1) axial-vector current and $K^\mu$
is the winding-number current
$$
K^\mu~ =~ {{g^2} \over {8\pi^2}}\epsilon_{\mu\alpha\beta\gamma}
Tr\bigg[-{{2ig} \over {3}}\epsilon_{ijk}A^i_\alpha
A^j_\beta A^k_\gamma + A^i_\alpha \partial_\beta A^i_\gamma \bigg]
\auto\label{wind}
$$

If an infinite momentum hadron has a ``cloud of topological glue'' around
it, the relation between topological winding number and vacuum production of
axial quark-antiquark pairs implied by (\ref{ano}) leads us to expect
corresponding divergences from massless quark intermediate states in our
Regge limit analysis. These are the divergences that we find.
(In this sense, the anomalous interactions which appear specifically in the
Regge limit can be understood as the ``perturbative realization'' of
instanton interactions.) The simplest situation occurs in SU(2) gauge theory
and the result is analagous to the solution\cite{man} of the two-dimensional
Schwinger model in the presence of the anomaly. That is, there is a vacuum
(``winding number'') condensate, that appears in our formalism via the
factorization of infra-red divergences, and {\it no scattering.} For SU(3)
gauge theory the high-energy Pomeron can be built up perturbatively on top
of an SU(2) vacuum condensate as follows.

With the SU(3) gauge symmetry broken to SU(2), our first approximation to
Pomeron exchange in hadron scattering is the Regge pole exchange
illustrated in Fig.~5.1. Perturbative exchange of the SU(2) singlet
reggeized gluon (the circle indicates the reggeization) between quarks
(represented by straight lines) is accompanied by a shadow process -
represented by the dashed lines. The shadow process is the effect of
massless SU(2) gluon exchanges, interacting via massless quark loops, which
produce the infra-red divergence that is absorbed onto the definition of the
states. We have used three dashed lines because the simplest divergent gluon
configuration involved\cite{arw1} is three gluons in a color-zero
combination, corresponding (loosely) to the three gluon component of the SU(2)
winding-number current. (The two gluon component occurs in combination with
the even signature octet reggeon referred to above).

\begin{center}

\epsffile{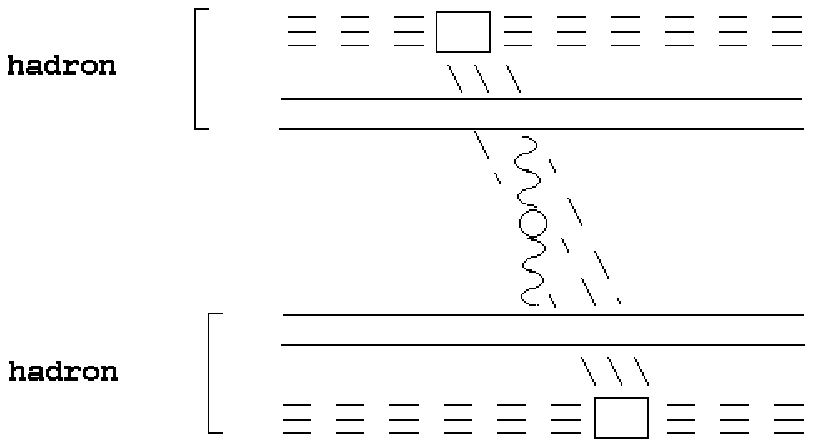}

Fig.~5.1 First Approximation to Hadron-Hadron Scattering

\end{center}

We believe that it should also be possible to view the shadow process as the
average effect of the exchange of winding number, in the topological sea of
each of the scattering hadrons, via instanton interactions. (As we emphasize
in \cite{arw2}, when $N_f=N_f^{max}$ there are no infra-red
renormalons in the massless theory. This implies that instanton interactions
are finite and represent the non-perturbative physics of the theory
completely.) We therefore suggest a space-time, or impact parameter picture,
of hadron-hadron scattering via Pomeron exchange as illustrated in Fig.~5.2.
We picture a hadron at rest as ``quarks in a bag''. At large momentum the
hadron expands in impact parameter space by virtue of the expansion of the
surface of the bag. This is approximated by a classical large impact
parameter field (essentially the ``wee-parton'' component of a hadron) which
we can think of as involving a sum over winding number topologies.

\begin{center}

\epsffile{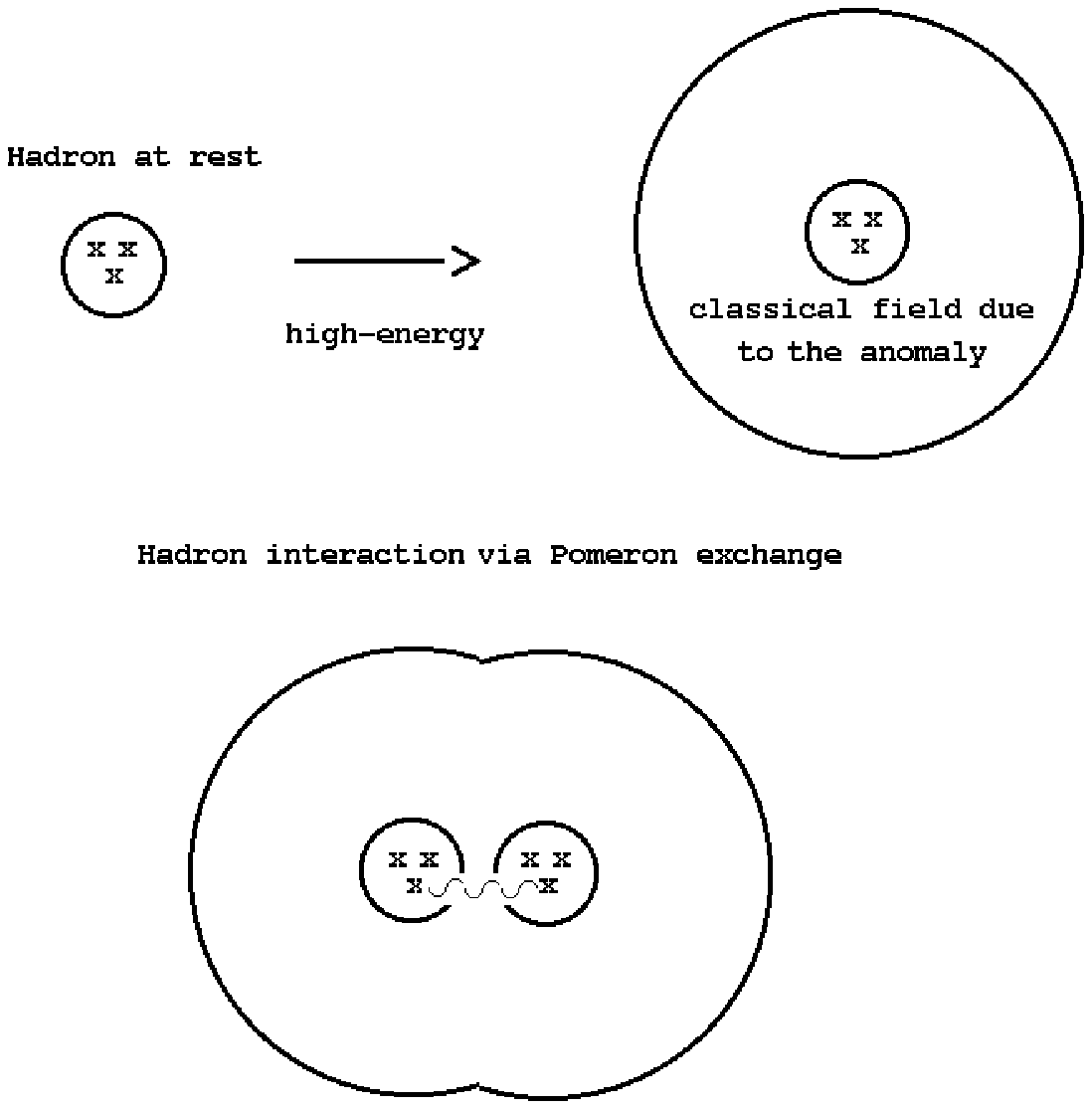}

Fig.~5.2 Impact parameter picture of high-energy hadron scattering as
perturbative gluon exchange within overlapping topological fields.

\end{center}

It is clear from Fig.~5.2 that hadron scattering at large impact parameter will
involve perturbative interactions taking place in the background of the
wee-parton fields. However, Fig.~5.1 implies that the condensate also
represents a wee-parton contribution to what is exchanged. (Indeed to obtain
an even signature Pomeron from gluon exchange, this must be the case.) We think
of this as a sum over (instanton) winding-number exchanges coupling the
relevant components of the hadron classical fields. If we anticipate that
Fig.~5.2 retains some validity after the restoration of SU(3) symmetry, it
is clear also that a transfer of SU(3) color in the perturbative
interaction can be negated by an (instanton) interaction between the
wee-parton fields which carries compensating color. We discuss this point
further at the end of the next Section.

The complete Pomeron is built up by multiple massive gluon exchanges and
interactions, which also couple to the condensate, as illustrated in Fig.~5.3
{\it For the restoration of the SU(3) symmetry it is crucial that
both the condensate and the reggeized gluon are SU(2) singlets but are
octets under SU(3) color}. The combination of the condensate and the gluon
have an SU(3) singlet projection and so form the SU(3) Pomeron. As the full
symmetry is restored, the massive gluon becomes massless, the gluons forming
the condensate become dynamical, and a Pomeron with unit intercept, the
Critical Pomeron, is obtained. In the process the singlet gluon also becomes
an important part of a hadron, combining with the condensate gluon
configuration to give an SU(3) invariant gluon (or Pomeron) contribution. In
this way a hadron regains it's conventional structure, becoming
predominantly an SU(3) singlet quark state, combined with a singlet
combination of wee-parton gluons. (As we noted above, the SU(2) singlet
component of the SU(3) symmetric octet two gluon reggeon\cite{bar} also
plays a crucial role in the formation of the SU(3) Pomeron).

\begin{center}

\epsffile{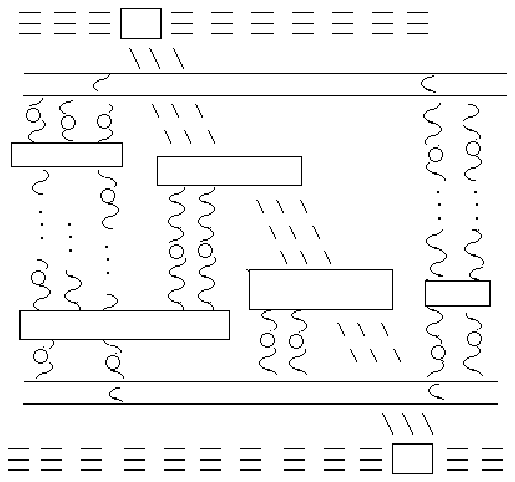}

Fig.~5.3 The Full Set of Pomeron Graphs

\end{center}

Finally we note that in a normal perturbative (small impact parameter)
interaction there can also be an instanton interaction between the
wee-parton fields. As the SU(3) symmetry is restored, a gluon interaction
generating the Pomeron must be constrained to carry the same color as the
wee--parton interaction, while in a perturbative gluon interaction this need
not be the case. As we observed in the Introduction this can be viewed as a
simple origin of the factor of $1/8$th in the relative cross--sections.

\mainhead{6. POMERON FIELD THEORY}

The idea that the strong interaction at high-energy is decribed by a
self-interacting Regge pole - the Pomeron - has a long history. In \cite{arw2}
it is argued that unitarity actually requires that asymptotically rising total
cross-sections be described by the strongly coupled Critical Pomeron.
The effective RFT near the Critical Pomeron fixed-point is a simple triple
Pomeron theory with the lagrangian
$$
{\cal L}~=~{1\over 2}\bar\psi
\centerover{${\partial\over \partial y}$}{$\leftrightarrow$}
\psi-\alpha'_0\nabla\bar\psi\nabla\psi + \Delta_0\bar\psi\psi -
{1\over 2}ir_0\left[\bar\psi\psi^2 + \bar\psi^2\psi\right],
\auto\label{7.19}
$$
where $\bar\psi$ and $\psi$ are, respectively, field operators creating and
destroying Pomerons in impact parameter and rapidity space. $r_0$ is the
triple Pomeron coupling and ${\alpha}_0'$ is the slope. For $\Delta_0 \equiv
(1-\alpha_{\spom}(0)) \gg 0$, the perturbation expansion defines the theory.
The critical behavior occurs at some $\Delta_0 = \Delta_{0C}$ and it can be
shown that if there is {\it no transverse momentum cut-off} then
$\Delta_{0C} = -\infty$. With a cut-off $\Lambda_\perp$, we have
$$
\Delta_{0C} = {r_O^2\over \alpha_0^\prime} \ln
{r_0^2\over\alpha_0\Lambda_\perp}
+ 0(r_0^2) ~~~< 0~.\auto\label{int}
$$
To define a (cut-off) theory with $\Delta_0 \ll \Delta_{0C}$, we look
for a classical field configuration minimizing the ``potential''
$$
V(\bar\psi, \psi) = \Delta_0\bar\psi\psi + {ir_0\over2} (\bar\psi^2\psi +
\bar\psi\psi^2).\auto\label{8.2}
$$
The relevant stationary point is
$$
\psi~ =~ \bar\psi~ =~ {2i\Delta_0\over 3r_0}
\auto\label{8.7}
$$
and the new potential is
$$
\tilde V(\bar\psi,\psi) = {-\Delta_0\over 3} \bar\psi\psi - {\Delta_0\over 3}
\bar\psi^2 - {\Delta_0\over 3} \psi^2 + {ir_0\over 2} (\bar\psi^2\psi +
\bar\psi\psi^2)\auto\label{8.8}
$$
giving a perturbation expansion that is stable for $\Delta_0~\ll ~0$, i.e.
$\alpha_0~\gg~1$. The $\bar\psi^2$ and $\psi^2$ terms are a direct
reflection of the presence of a Pomeron condensate.

When the RFT graphs describing this Super-Critical theory are
constructed\cite{arw2} and analysed via reggeon unitarity, we find that
there is an odd-signature trajectory degenerate with the Pomeron (both have
intercept below one). The odd-signature reggeon couples pairwise to the
Pomeron. In detail we find that the reggeon diagrams of the Super-Critical
Pomeron theory have just the structure of those described in the last
Section, i.e. the reggeon diagrams of QCD with the gauge symmetry broken
from SU(3) to SU(2). The Pomeron condensate is identified with the
reggeon (wee parton) condensate arising from the infra-red divergences.

If we assume that, in the neighborhood of the critical point, the parameters
of QCD map straightforwardly  on to those of RFT, a number of conclusions
follow. Firstly, although we constructed the Super-Critical phase by using
the Higgs mechanism to ``break the gauge symmetry'', this is not necessary.
The true characterization of the super-critical phase is an exchange
degenerate Pomeron together with a Pomeron condensate. In general
$\Lambda_\perp$ is a relevant RFT parameter at the critical point, in addition
to $\Delta_0=(1-{\alpha}_0)$. The analysis of the previous Section shows that
in the particular case $N_f=N_f^{max}$ the critical point is at
$\Lambda_\perp =\infty$. Assuming, as follows from(\ref{int}), that
$\Delta_0~<~\Delta_{0C}$ corresponds to $\Lambda_\perp < \Lambda_{\perp C}$, it
is clear that the Super-Critical phase can also be reached without
introducing an additional ``Higgs parameter''. The RFT
analysis implies that when $N_f = N_f^{max}$ the Super-Critical Pomeron
occurs when $\Lambda_\perp < \infty$. When $N_f ~<~N_f^{max}$, the critical
point is at a finite value of the transverse momentum cut-off and so the {\it
Super-Critical phase can be introduced by taking $\Lambda_\perp <
\Lambda_{\perp C}$. }

Let us consider, therefore, how the Super-Critical Pomeron can occur in an
SU(3) invariant manner. It might appear that a simple SU(3) invariant version
of our discussion would involve the usual SU(3) winding-number current,
in which the ${\epsilon}_{ijk}$ in (\ref{wind}) are replaced by the SU(3)
structure constants. However, since this is a color zero operator, it is
clear that the Pomeron could not appear as gluon exchange in the background
of a related condensate. Instead the Super-Critical Pomeron must be
associated with interactions of the kind represented by Fig.~5.1 and
Fig.~5.2, but with the color of the exchanged gluon and that of the
classical hadron fields summed and averaged over. In fact it is well known
that instantons are essentially an SU(2) phenomenon\cite{bcgw} and that
the simplest topology SU(3) instantons are always confined within
an SU(2) subgroup. Our present discussion suggests that dynamical effects of
the fermion anomaly must be accounted for before instanton parameters,
corresponding to group rotation of the sub-group, are summed over to obtain
an SU(3) invariant result. This implies, in particular, that a
Super-Critical hadron is an SU(3) triplet in a non-trivial way. That is, the
quark component of the hadron is in an octet state corresponding to the
SU(3) generator that leaves an initial instanton subgroup unchanged. The
``topological glue component'' resulting from instanton interactions within
that subgroup is then necessary to obtain an overall singlet. The color
direction of the octet is summed over as part of the process of summing
over instanton parameters.

In the Super-Critical phase, therefore, diffractive scattering can be
calculated using a massive gluon. The presence of a transverse momentum cut-off
is crucial in that the mass is generated dynamically by taking the cut-off
{\it below} the critical value. Physically this will atomatically be the
case if the critical cut-off is above presently accessible transverse
momenta. It is interesting to ask what physically sets the scale of this
mass. We anticipate that $\Lambda_{\perp C}$ is a function of
$N_f^{max}-N_f$, but from (\ref{int}) we see that the cut-off enters only
logarithmically in determining how close to criticality the Pomeron is. As a
result the gluon mass scale is mainly set by
$$
m_g^2~~=~~{\Delta_{0C}~-~\Delta_0 \over \alpha_0'}~~\sim~~ {r_0^2
\over {\alpha_0'}^2}~~
\auto\label{sca}
$$
We anticipate that $r_0~\sim~1/m_{\pi}$ and
$\alpha_0'~\sim~1/ \lambda^2_{QCD}$, in which case
$$
m_g~~\sim~~{\lambda^2_{QCD} \over m_{\pi} }
$$

In the physical world, as currently explored, we have both $N_f<N_f^{max}$
and of course, simply because of the limitation of finite energy,
$\Lambda_\perp <\infty$. It also seems to be established that the ``soft
Pomeron'' is well described phenomenologically\cite{dl} by an (unstable)
Super-Critical Pomeron expansion with a bare Pomeron intercept
$\alpha_{\spom}(0)>1$. Clearly we can consistently assume that we are in the
Super-Critical phase. In the RFT framework, therefore, the Super-Critical
expansion involving a Pomeron condensate is really the most appropriate way
to describe the Pomeron. Although we should note that RFT graphs describe
idealised (asymptotic) physical processes in which finite-energy kinematic
limitations on sub-energies and transverse momenta are not imposed.
Nevertheless, as we have discussed, the Super-Critical graphs are related
(albeit in a complicated way) to perturbative QCD processes. Also the gluon
mass can be regarded as induced by the very large transverse momentum region
(making it sensitive to as yet unexplored higher momentum scales).
Consequently the Super-Critical formalism is very likely to be the most
appropriate for discussing diffractive hard-scattering processes. This
clearly is the point of view we advocate as providing the correct
interpretation of the BH picture for deep-inelastic scattering.

We should emphasize that if the theory is truly in the Super-Critical phase
then RFT implies that the total cross--section will ultimately fall
asymptotically. This can not be consistent with the validity of QCD
perturbation theory at large transverse momentum. As we already remarked in
the Introduction, we believe that, for full consistency the additional
fermion sector, needed to obtain the Critical Pomeron with no transverse
momentum cut-off, must eventually enter the theory.

\mainhead{ 7. DEEP-INELASTIC SCATTERING}

Let us return now to the QCD Pomeron in the idealised situation where we
have $N_f~=~N_f^{max}$ and the gauge symmetry is explicitly broken to SU(2).
Consider deep-inelastic scattering of a hadron and a photon carrying
large (spacelike) $Q^2$. Let us compare the forward amplitude giving the
total cross-section with the diffractive cross-section.
In the calculation of the total cross-section, the mass we have
introduced has little impact. Asymptotic freedom exposes the elementary
quark-antiquark coupling at large $Q^2$ and we can suppose also that the
kinematics of the final state justify selecting Fig.~1.2 as the dominant
perturbative process in the forward amplitude. (As we discussed earlier,
this a subtle point since Fig.~1.2 successfully describes far more of the
total cross-section than it should.) The gluon involved can be either
massless or massive and as the SU(3) gauge symmetry is restored the
distinction becomes irrelevant. (Since we are not discussing diffractive
scattering, there is no appearance of the condensate). We therefore obtain
the usual full QCD contribution of Fig.~1.2 to the total cross-section.

For the diffractive cross-section corresponding to the same quark-antiquark
contribution to the final state we obtain Fig.~7.1.

\begin{center}

\epsffile{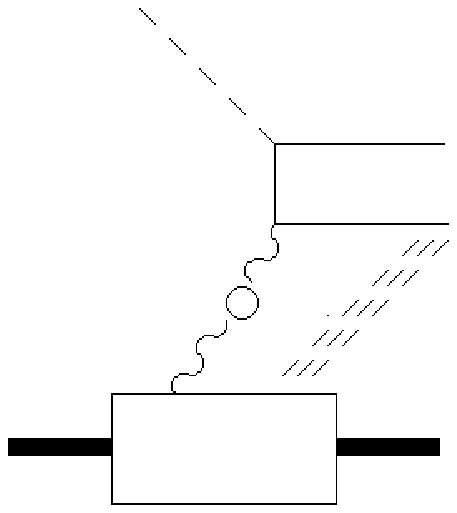}

Fig.~7.1 Diffractive Deep-Inelastic Scattering

\end{center}

\noindent  We could show the photon
as having a ``topological sea'' but this should, of course, be irrelevant at
large $Q^2$. Therefore we have shown the exchanged infra-red divergent
gluons as absorbed by the parton final state without any rescattering on the
corresponding component of the photon. Large $Q^2$ again determines that the
photon couples perturbatively, as we have shown in Fig.~7.1. Now, since the
Pomeron is involved, it is necessarily the massive SU(2) singlet reggeized
gluon that appears. The box in the lower half of Fig.~7.1 contains the full
set of Pomeron graphs having, essentially, the structure illustrated in
Fig.~5.3.

The impact parameter picture corresponding to Fig.~7.1 is shown in Fig.~7.2.
The elementary photon enters the center of the hadron and produces an
additional quark pair. There is a perturbative gluon exchange before the
classical field component of the hadron breaks apart in a manner that, as the
SU(3) symmetry is restored, compensates for the color exchange.

\begin{center}

\epsffile{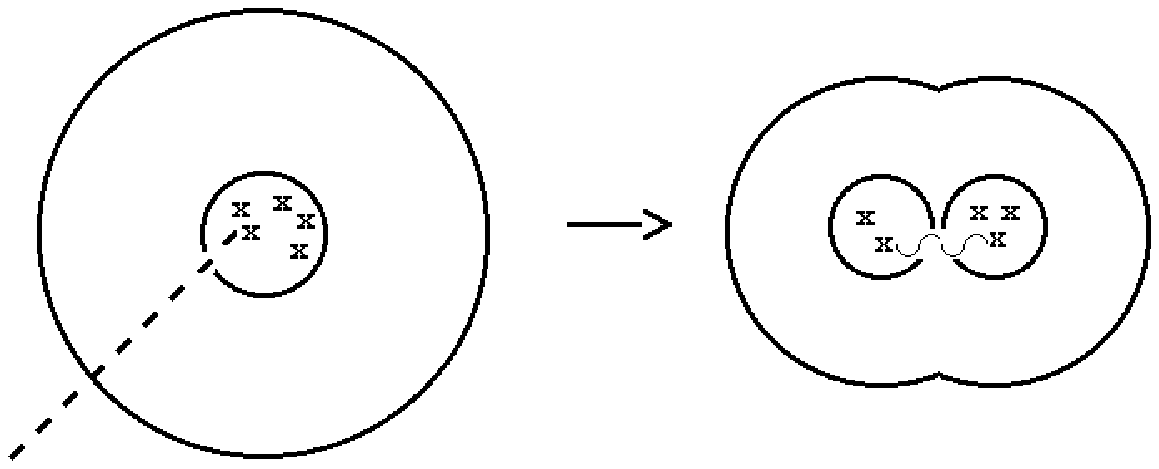}

Fig.~7.2 Impact Parameter Picture of Diffractive Deep-Inelastic Scattering

\end{center}

There are two crucial ingredients that lead us to the BH result for the
scaling violations - as the SU(3) symmetry is restored. The first is that
{\it the condensate disappears smoothly.} The second is that {\it asymptotic
freedom continues to select the perturbative photon-gluon scattering
process} - in the region of the cross-section that is insensitive to
$m_g^2$. We are, of course, effectively assuming that the deep-inelastic,
diffractive, and symmetry restoration limits all commute for this part of
the cross-section.

To discuss the $x$-dependence, let us first make the usual BFKL assumption that
the small value of $\alpha_s$ generated by the deep-inelastic limit justifies
using {\it the leading log[$1/ \xi$] approximation}. Taking the SU(3) symmetry
limit and considering the full cross-section for the diffractive excitation
of the proton we obtain a simple result. The condensate disappears and,
because the reggeization of the gluon is included in the Pomeron, the
leading-order BFKL equation appears with the SU(2) singlet gluon still
coupled to the quark/antiquark pair. This is illustrated in Fig.~7.3, where
$K$ denotes the BFKL kernel and we have used a simple loop to represent the
leading log reggeization. The solution gives the small-x (singular)
structure function $g(\xi)$, as required. According to our formalism
therefore, it is consistent that the BFKL Pomeron gives the behaviour of
parton distributions at small-x, even though it does not describe the physical
Pomeron that is responsible for a rapidity gap.

\begin{center}

\epsffile{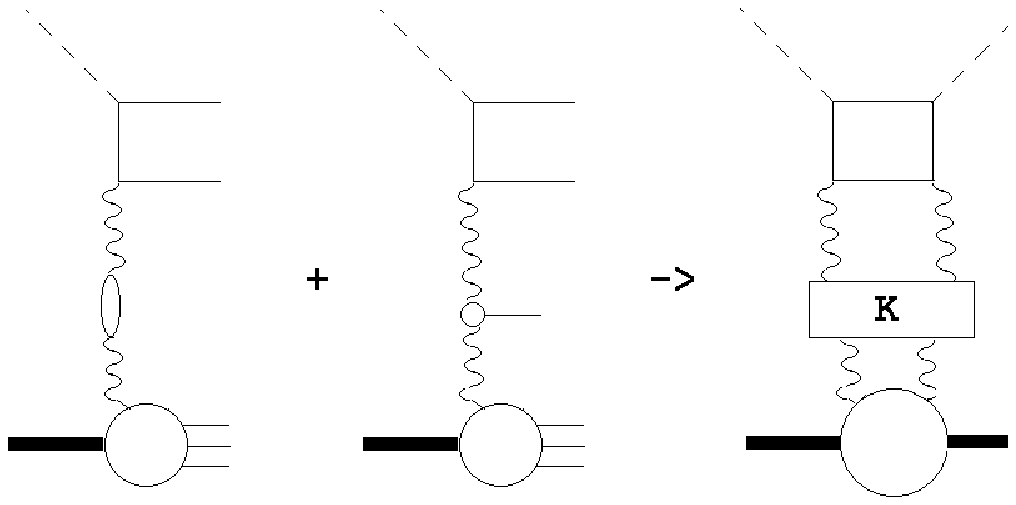}

Fig.~7.3 The BFKL equation for the Diffractive Cross-section

\end{center}

In the leading-log approximation, we obtain all the
ingredients of the BH cross-section. The normalization is clearly $1/8$th
since the cross-section is specifically given by a single gluon, i.e
the gluon that is an SU(2) singlet under the symmetry-breaking procedure,
and we could have chosen any of the eight gluons to play this role. To
extend the argument beyond leading-logs, we need to discuss the contribution
of the symmetric octet two-gluon reggeon\cite{bar} to the Super-Critical
Pomeron and the eventual emergence of the Critical Pomeron. The
self-consistency properties of the Critical Pomeron will again give the scaling
property, but with an undetermined relative normalization.

Note that for Fig.~7.1 to make sense it is crucial that the quark-antiquark
state can combine with the SU(2) singlet condensate to make a hadronic final
state. As the SU(3) symmetry is restored, this requires that it be an SU(3)
octet, which of course it is. In this sense it is vital that we consider
Fig.~1.2 throughout the full kinematic regime. The normal formulation of the
parton model within QCD requires that we separate kinematic configurations
in which either the quark or antiquark is relatively soft and sum over
related processes to introduce parton distributions. In this case it would
not be consistent to introduce the condensate as we have done. In general,
for a short-distance parton process to couple to the Pomeron in analogy with
our present discussion, we believe it is necessary that the parton state be
a quark state that can similarly be converted to a hadronic final state. (It
is possible that the final state could also be a massive SU(2) singlet gluon
state, before the gauge symmetry is restored to SU(3). This relates to the
issue of whether there are glueballs in the physical spectrum and for the
present we will assume that this is not the case.)

For the BH model to be applicable to QCD with just the normal number of quarks,
it is very clear from the discussion of this and the previous Section what
we must assume to be the case. The effective transverse momentum cut-off
imposed by current energies must be below the critical cut-off, so that the
Super-Critical Pomeron phase is realised. In addition an approximate form of
asymptotic freedom must still be valid to justify using deep-inelastic
scattering to expose single gluon exchange. (Of course, this must also be
the case to justify the general success of perturbative QCD!!)

If the diffractive deep-inelastic scaling property is to survive
asymptotically, we must have full asymptotic freedom combined with the
proximity of the Super-Critical phase. This implies we must have the
conditions discussed in Section 4, i.e. a higher-mass quark sector must
eventually enter the theory.

\mainhead{8. DIFFRACTIVE HARD SCATTERING AT THE TEVATRON}

We now consider the case of Super-Critical diffractive $W$ production in
hadron-hadron scattering. (We no longer distinguish the explicit breaking of
the gauge symmetry from the case of unbroken QCD - in which
we simply average over the SU(2) direction.) Assuming the large momenta
involved exposes a parton production process, our previous discussion
implies there must be a quark-antiquark pair in the final state. Therefore
we should consider a process of the form shown in Fig.~8.1
Since the large-momentum quark pair is most likely to be produced
by a gluon we expect the leading-order diffractive $W$ production to be via
the ``gluon-gluon fusion'' process illustrated in Fig.~8.2.

We must, however, qualify what is meant by gluon-gluon fusion in our case.
In analogy with our discussion of photon-gluon fusion earlier, predominantly
one of the quark/anti-quark pair will be relatively soft. The $W$ must be
produced by the hard quark(anti-quark) and so the correct parton model
description is the gluon-quark (gluon-antiquark) process of Fig.~8.3 with
the soft anti-quark (quark) production absorbed in the structure function.

\begin{center}

\epsffile{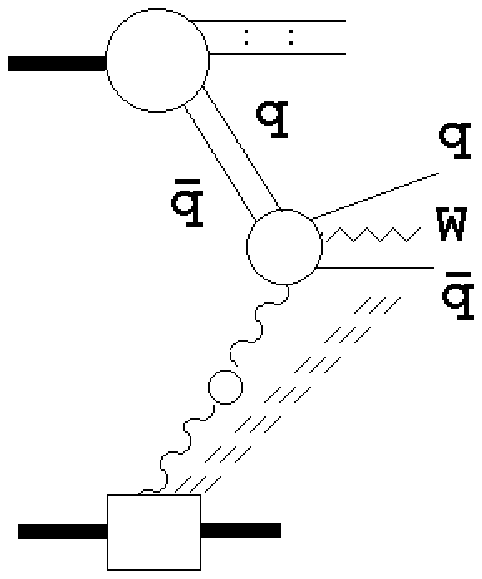}

Fig.~8.1 Diffractive W Production

\end{center}

\begin{center}

\epsffile{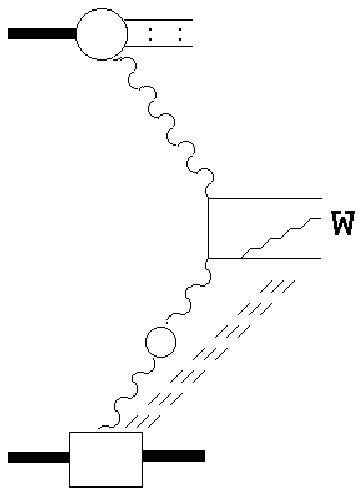}

Fig.~8.2 Diffractive W Production via Gluon-Gluon Fusion

\end{center}

\begin{center}

\epsffile{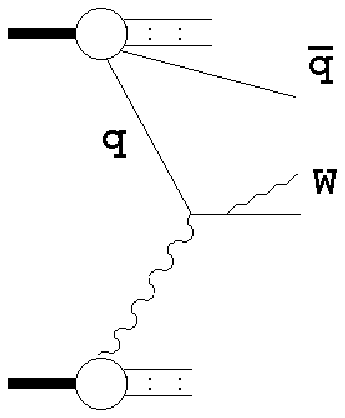}

Fig.~8.3 W Production via Quark-Gluon Scattering

\end{center}

\noindent As we have emphasized, the role of the additional soft ``parton''
is crucial in the diffractive process. Therefore the necessity for it's
presence must be allowed for when we compute the full parton scattering
process that we wish to multiply by $1/8$ th, just as the full
deep-inelastic cross-section was used to estimate the cross-section for
Fig.~1.2. The effect is that we must, in principle at least, extract from
all quark and anti-quark parton distributions (including valence quarks and
antiquarks) the contribution due to an intermediate gluon.

Given the increase of the gluon distribution at small-x, the total contribution
of gluon-gluon scattering to all processes of the form of Fig.~8.3, {\it
including both valence and sea quarks and antiquarks}, could easily be as
high as $10\%$ of the total cross-section. In this case we would expect the
diffractive cross-section to be of the order of $1\%$ of the total. This is
certainly compatible with, and perhaps even suggested by, the CDF analysis
of rapidlty-gap $W$ production\cite{cdf}.

As we mentioned above, there is, in our view, a problem with the interpretation
of the CDF data in terms of the Ingelman-Schlein Pomeron hard scattering
formalism - in which the Pomeron is assumed to have a hadronic structure
function. Independently of whether quarks or gluons dominate the Pomeron
structure function the Pomeron predominantly scatters of a quark (or
antiquark), as illustrated in Fig.~8.4.

\begin{center}

\epsffile{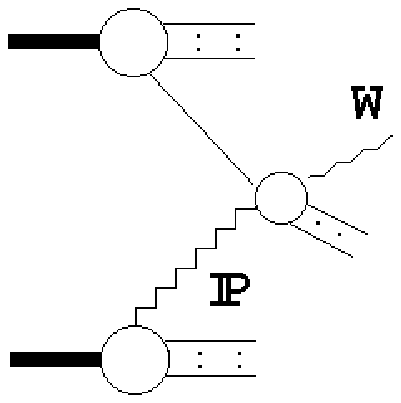}

Fig.~8.4 W Production via Pomeron-Quark Scattering

\end{center}

\noindent If conventional parton distributions are assumed for the quark
scattering off the Pomeron, clearly valence quarks (and antiquarks) will
dominate. As a result there should be a strong correlation (in
proton-antiproton scattering) between the charge of the $W$ and where it
appears on the rapidity axis relative to the rapidity-gap. In fact there is no
evidence for such a correlation in the candidate gap events, leading CDF to
make the strong conclusion\cite{cdf} that they see no diffractive $W$
production. Since the multiplicity distribution that would extrapolate to
the gap cross-section looks identical for $W$ and jet production, this leads
to the even stronger conclusion that they see no diffractive di-jet
production.

It is clear that our formalism does not predict the charge/gap correlation
that the Ingelman-Schlein approach predicts. In our case, the requirement that
the final parton state be consistent with confinement implies that the $W$ is
produced from a quark-antiquark pair. This requires in particular that the
quark in Fig.~8.4 be accompanied by an antiquark, and in general that either
charge is equally likely in any diffractive configuration. The production of
the quark pair by a gluon also implies that the $W$ will be produced
preferentially close to the rapidity gap. This goes against the expectation
that the W will be produced predominantly on the opposite side of the
rapidity axis to the gap. Such a correlation is also looked for and
apparently not seen in CDF data. Since there are other features of the CDF
data for both $W$ and jet production that suggest diffractive production is
indeed present, we believe our picture provides a consistent and attractive
interpretation of what is seen.

\mainhead{9. COMMENTS AND CONCLUSIONS}

We have believed for some time that many fundamental problems are related to
understanding the Pomeron in QCD. It involves an intricate combination of
short-distance and long-distance dynamics whose solution is surely relevant
for any gauge theory. We hope that the striking nature of the scaling
behavior displayed in Fig.~1.1 will provoke a serious theoretical effort to
determine the implications for the Pomeron. Our work\cite{arw1} implies
these are significant. However, since establishing the picture we have
developed would include, in particular, a proof of (high-energy)
confinement, it is not too surprising that much remains to be done.
Nevertheless developing a simple, consistent, phenomenology should be
possible. For this, the connection that we have made with the Buchm\"uller and
Hebecker model\cite{bh} ought to be particularly valuable.

Finally we comment on the possibility that deep-inelastic diffractive
scaling actually reflects asymptotic properties of the Pomeron and what this
implies for new physics at higher energy scales. As we hope our discussion
demonstrates, the interplay between deep-inelastic and diffractive scaling
actually probes virtual properties of the theory that are related to mass
scales yet to be exposed as physical states. We have elaborated elsewhere
(e.g. in the summarizing chapter of \cite{arw1} and in \cite{arw3}) our
belief that the missing states that will ultimately be responsible for the
Critical behavior of the Pomeron are a higher color quark sector that is
responsible for dynamical symmetry breaking and CP violation in the
electroweak sector, as well as a major transformation in strong-interaction
physics! There may already be evidence for this physics in Cosmic Rays and
in the highest energy hard scattering processes at the Tevatron\cite{arw3}.

\vspace{1in}

\subhead{Acknowledgements}

I would like to thank J. Bartels, G. Bodwin, W. Buchm\"uller and A. Hebecker
for valuable discussions and information.

\end{document}